# Reducing Magnetic Resonance Image Spacing by Learning Without Ground-Truth

Kai Xuan[a], Liping Si[b], Lichi Zhang[a], Zhong Xue[c], Yining Jiao[a], Weiwu Yao[b],

Dinggang Shen[d,e], Dijia Wu[c], and Qian Wang[a,*]

[a]*Institute for Medical Imaging Technology, School of Biomedical Engineering, Shanghai Jiao Tong University, Shanghai 200000, China*
[b]*Department of Imaging, Tongren Hospital, Shanghai Jiao Tong University School of Medicine, Shanghai 200000, China*
[c]*Shanghai United Imaging Intelligence Co., Ltd., Shanghai 200000, China*
[d]*Department of Radiology and BRIC, University of North Carolina at Chapel Hill, Chapel Hill, NC 27599, USA*
[e]*Department of Brain and Cognitive Engineering, Korea University, Seoul 02841, South Korea*
*Corresponding author: e-mail: wang.qian@sjtu.edu.cn (Qian Wang)*

# ABSTRACT

High-quality magnetic resonance (MR) image, i.e., with near isotropic voxel spacing, is desirable in various scenarios of medical image analysis. However, many MR images are acquired using good in-plane resolution but large spacing between slices in clinical practice. In this work, we propose a novel deep-learning-based super-resolution algorithm to generate high-resolution (HR) MR images of small slice spacing from low-resolution (LR) inputs of large slice spacing. Notice that real HR images are needed in most existing deep-learning-based methods to supervise the training, but in clinical scenarios, usually they will not be acquired. Therefore, our unique goal herein is to design and train the super-resolution network without real HR ground-truth. Specifically, two-staged training is used in our method. In the first stage, HR images of reduced slice spacing are synthesized from real LR images using variational auto-encoder (VAE). Although these synthesized HR images of reduced slice spacing are as realistic as possible, they may still suffer from unexpected morphing induced by VAE, implying that the synthesized HR images cannot be paired with the real LR images in terms of anatomical structure details. In the second stage, we degrade the synthesized HR images to generate corresponding LR-HR image pairs and train a super-resolution network based on these synthesized pairs. The underlying mechanism is that such a super-resolution network is less vulnerable to anatomical variability. Experiments on knee MR images successfully demonstrate the effectiveness of our proposed solution to reduce the slice spacing for better rendering.

*Index Terms: Generative adversarial network, magnetic resonance imaging, super-resolution, variational auto-encoder.*

# 1 INTRODUCTION

Magnetic resonance (MR) images are widely used in clinics. But due to the issues such as signal-noise ratio and scanning time, many of them are acquired with high in-plane resolution but large spacing between slices (2D acquisition), although images with near isotropic voxel spacing are highly desirable especially for 3D rendering.

It is appealing to enhance the resolutions of the acquired MR images in many clinical scenarios [1]. Among them, to reduce the slice spacing is a focus. Because compared to the in-plane resolution, the spacing between slices is often much larger. If the slice spacing can be effectively reduced, i.e., to the comparable level with the in-plane resolution without introducing many artifacts, a quasi-isotropic rendering of the acquired image can be attained.

To this end, we aim to investigate the problem of reducing the slice spacing of MR images in this work. We have particularly developed a learning-based super-resolution approach. For convenience, we will term the acquired MR image (3D volume) of large slice spacing as the *low-resolution (LR) image*, and the image with small slice spacing as *high-resolution (HR) image*, respectively. Examples of the LR and HR images can be found in Fig. 1.

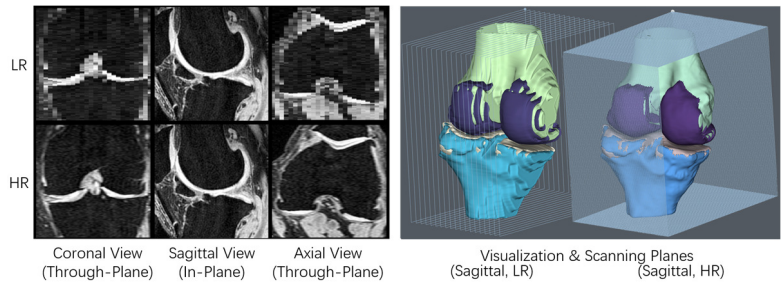

*Fig. 1. Example of LR and HR knee MR images scanned in sagittal views (left), and the 3D renderings of the scanned images based on their bone/cartilage segmentation (right). The through-plane resolution is much lower than those of the in-plane directions. The renderings of the anatomic structures are thus much better in HR image.*

Deep learning has already become a major tool to attain image super-resolution. A typical deep learning solution to super-resolution usually relies on three components, i.e., a large dataset containing paired HR and LR images, a generator to invert the image degradation (i.e., from HR to LR) process, and an optimization scheme to train the generator efficiently. A lot of previous works were published to better model the degradation process and to better design or train the generator underlying the prior knowledge of degradation, which will be surveyed below.

A common bottleneck in medical image super-resolution is the lack of real HR images. That is, paired LR/HR images are not popularly available. Even one can acquire HR and generate corresponding LR images, HR images are not easy to collect in terms of big training data. Thus, many conventional super-resolution methods cannot apply, since most of them need full voxel-to-voxel supervision. In this way, we will address this problem of reducing slice spacing by deep-learning-based super-resolution, while no real HR images are available to supervise the training of our deep networks.

## 1.1 MODEL-BASED SUPER-RESOLUTION

The objective of super-resolution is to reconstruct HR images from corresponding low-quality inputs. The technique can be generally categorized to model-based and learning-based approaches.

The model-based methods assume a priori known image degradation model and aim at finding an optimal solution to the HR image by optimizing the HR estimation with an objective function and the actual degraded acquisition. Since the inverse modeling of the degradation process is ill-posed, the objective function usually consists of two terms. The fidelity term requires the HR estimation after being manually corrupted to be close to the actual LR image; the regularization term excludes many ambiguities in searching the solution space.

The regularization term employs strong prior knowledge and plays an important role to deliver high-quality super-resolution results. Among them, total variation (TV) measurement is widely adopted [2]. Additionally, formulating the low-rank nature of brain MR images, Shi *et al.* [3] successfully improved the reconstruction quality of HR images with the low-rank total variation (LRTV) regularization. While the regularization term is usually hand-crafted, Chen *et al.* [2] proposed to integrate a priori knowledge to learn the regularization term in a data-driven fashion.

There are also some researches in other areas of the problem than the regularization terms. Tourbier *et al.* [4] adopted variable hyper-parameters for the trade-off between the fidelity loss and the regularization term when solving the problem of fetal brain reconstruction. Hatvani *et al.* [5] introduced tensor factorization to speed up the optimization process. Instead of integrating prior knowledge with the regularization term in the loss function, Dalca *et al.* [6] constrained the solution space to HR images that can be generated with Gaussian Mixture Models (GMMs). Instead of directly finding the HR estimation, they optimized the parameters of GMMs.

## 1.2 LEARNING-BASED SUPER-RESOLUTION

Recently, learning-based super-resolution algorithms have become state-of-the-art and widely used in a variety of applications. Different from the approaches relying on pre-defined models, the learning-based methods alleviate the demands on prior knowledge and implement super-resolution in a more flexible data-driven way. Sparse coding [7] and random forest [8] are popularly adopted, especially to model the LR-HR mapping in the patch-wise manner. For example, they are successfully used by Zhang *et al.* [9] to reconstruct isotropic brain MR images from high-thickness acquisitions.

Enabled by the high-performance computing facility and huge amount of training data, deep learning has become a powerful tool in super-resolution. Dong *et al.* [10] proposed to perform end-to-end training of the fully convolutional network (FCN) [11] to reconstruct natural HR images from LR inputs; Jiang *et al.* [12] proposed a wider and deeper yet efficient neural network for better performance; Yang *et al.* [13] further introduced channel-attention and graph convolution network (GCN) for better feature extraction and correlation modeling. Deep-learning-based super-resolution is not only accurate, but also robust in wild-variety of tasks, such as enhancing images for better person re-identification [14] or recognizing faces from extremely low-quality data [15], and it can also be used in depth images [16].

In medical imaging, Chaudhari *et al.* [17] implemented a 3D convolutional neural network (CNN) and outperformed conventional super-resolution methods significantly. More recently, Chen *et al.* [18], employing 3D densely connected generator [19] and adversarial training, produced high-quality MR images that are almost indistinguishable from real HR ones.

When applying deep learning to medical image super-resolution, the lack of training data is inevitable in many circumstances. A typical framework for learning-based super-resolution approaches can be found in Fig. 2 (a). With millions of parameters, a deep generator relies on a large dataset containing paired LR-HR images. We note that most existing works tend to manually degrade the HR images to synthesize the LR images [20]. However, such paired data are not always available, since the HR medical images can often be impossible to acquire for supervising the training process.

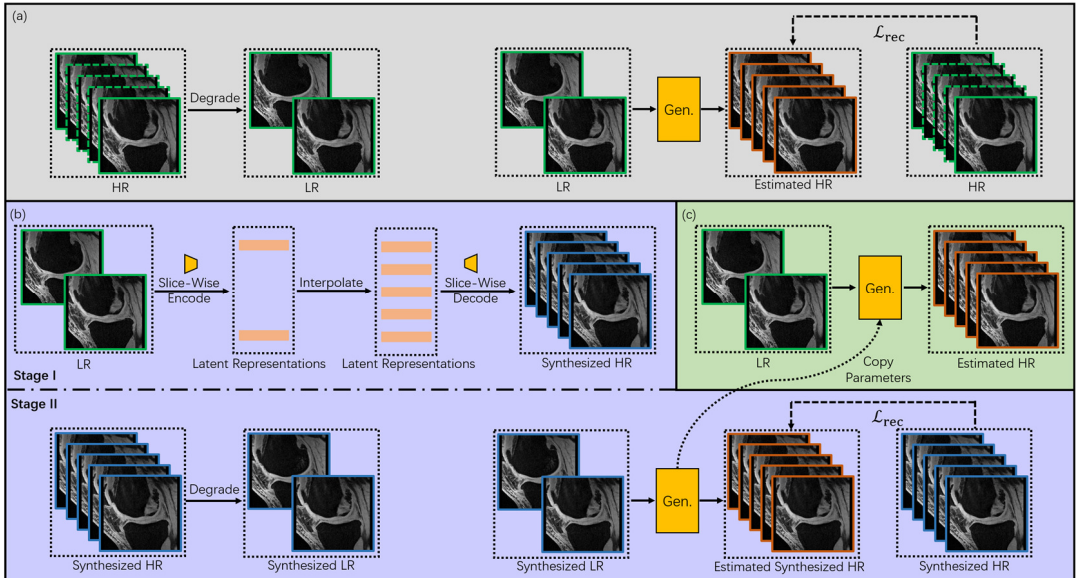

*Fig. 2. Illustration of our learning-based super-resolution framework. (a) In conventional learning-based super-resolution, real HR volumes are first degraded to LR volumes. Then a generator inverting the degradation process is trained, with the LR volumes as the inputs, and the HR volumes as the supervision to minimize the reconstruction loss $\mathcal{L}_{rec}$. (b) The training of our proposed framework consists of two stages. In Stage I, we train a 2D VAE on real slices from LR images and then synthesize HR volumes from real LR volumes by the VAE-based up-sampling. In Stage II, we first degrade the synthesized HR volumes to the synthesized LR ones and then a 3D super-resolution generator is trained on the synthesized LR-HR pairs. In this way, we avoid using real LR-HR pairs to train the generator as in conventional methods. (c) To test our proposed framework, a real LR volume is directly fed into the generator trained in (b), and the output is the estimation of the corresponding HR volume.*

## 1.3 SELF SUPER-RESOLUTION

Different from learning upon paired LR-HR images, self super-resolution can complete the task with LR inputs only. In natural images, the patches of similar patterns may recur frequently in various locations but with different scales. It is thus feasible to utilize the non-local strategy to identify patches with corresponding appearance of HR-scales for an LR patch. In this way, the paired LR-HR patches can be acquired within a single input LR image only as its *internal exemplars* [21]. While the quality of the training data is pivotal to the performance of super-resolution algorithms, Huang *et al.* [22] further expanded this strategy beyond translation to search for internal LR-HR exemplars and got comparable performance with respect to the results trained with *external exemplars* (i.e. LR-HR patches extracted from paired LR-HR images).

However, it is non-trivial to transfer similar ideas to medical images. Although imaging devices and protocols can vary a lot, for most clinical acquisitions the in-slice voxel spacing is distributed in a narrow range of physical units, which is unable to produce many internal exemplars for the organs and objects of interest especially compared to natural images. Alternatively, Zhao *et al.* [23] proposed to reduce slice spacing by training the enhanced deep self super-resolution network (EDSSR) within 2D in-plane slices. That is, by training a generator for super-resolution in 2D slices, one may apply the trained network for super-resolution along the through-plane direction and reduce the slice spacing. The resulted intermediate 3D HR estimates are further fused in the frequency domain for the final super-resolution result.

Currently deep-learning-based methods with full supervision (i.e., well-paired LR-HR training images) have demonstrated superior performance for MR image super-resolution [22]. If only the internal exemplars (i.e., paired LR-HR patches acquired from input LR images only) are used without the ground-truth of HR images, the produced MR images still fall behind real HR acquisitions by significant quality gaps [23,24]. Taking EDSSR [23] for example, the drawbacks may lie on both the self-extracted training exemplars (i.e., all training examples are extracted from axial views while the expectation is to restore coronal and sagittal HR views) and the generator (i.e., by modeling 2D LR-HR mapping [25] and applying it to 3D volumes).

## 1.4 OVERVIEW OF THE PROPOSED METHOD

In this work, we propose a deep-learning-based super-resolution framework to reduce the slice spacing in MR images. Our work is unique as the real HR images are not needed to supervise the training of the network. And we expect that our framework can deliver comparable super-resolution performance with respect to the case of real HR supervision. An overview of our method is available in Fig. 2 (b) for training and Fig. 2 (c) for testing. Particularly, there are two stages to train the super-resolution generator.

- In Stage I, we synthesize HR volumes given only LR volumes.
    - First, we optimize a variational auto-encoder (VAE) with 2D slices extracted from LR images.
    - Then, HR images of reduced slice spacing are synthesized from real LR images, by applying the trained VAE to 2D slices in the real LR images. This is done by the following steps:
        1. Mapping 2D MR slices into a compactly distributed latent space;
        2. Interpolating on the latent representations;
        3. Decoding the input slices and the interpolated representations.
- In Stage II, we use the synthesized HR images (instead of real HR images that are not available in practice) to supervise the training of a deep learning model.
    - We degrade the synthesized HR images to generate corresponding LR images.
    - A fully supervised method is trained on the LR-HR pairs to model the inverse of the degradation process and obtain the optimal HR estimation.

In testing, the trained generator can be directly used for a new LR image of large slice spacing, from which the corresponding HR output of reduced spacing can be acquired.

The advantage of the proposed deep-learning-based method is that it uniquely relies on LR images only. Without referring to any real HR data to supervise the training, we still can achieve comparable performance with state-of-the-art fully supervised super-resolution methods. In addition, we have substantially improved the encoder-decoder architecture by imposing novel local linearity constraint in VAE, which encourages the 2D slices to compactly distribute in the latent space. In this way, the synthesized 3D HR image can be as realistic as possible, which helps train the generator for super-resolution. Experiments on knee MR images successfully demonstrate the effectiveness of our proposed solution.

## 2 METHODS

We will introduce the details of our deep-learning-based super-resolution method in this section. First, in Section 2.1, we retrospect the mathematic modeling of super-resolution and derive our solution when real HR supervision is not available. Then, we present the details of the training scheme of the proposed method in two stages. The training process of our method is illustrated in Fig. 2(b), with the testing process in Fig. 2(c).

### 2.1 LEARNING-BASED SUPER-RESOLUTION

With a pair of the HR image $X$ and the corresponding LR image $Y$, the image mapping can be expressed as $Y = T \cdot X + N$, where $T$ approximates down-sampling or degradation process, and $N$ is the additive noise. Super-resolution is then defined as the inverse process of obtaining the optimal HR estimation $\hat{X}$ from the corresponding LR image $Y$.

With priors of the degradation process, the model-based approaches can directly estimate $X$ from $Y$ by minimizing the following cost function

$$\hat{X}^* = \arg\min_{\hat{X}} \|T \cdot \hat{X} - Y\| + \lambda R(\hat{X}), \tag{1}$$

where $\|\cdot\|$ is the reconstruction fidelity (e.g., L2-norm or L1-norm) encouraging the estimated HR image to be similar with the observation $Y$ after corruption, $R$ is the regularization term to the solution space, and $\lambda$ is a hyper-parameter for trading-off between the two terms.

Different from the model-based approaches, the learning-based approaches aim to find the mapping from LR to HR images in the data-driven way. The training process often requires a large dataset of paired $X$ and $Y$. Instead of directly estimating $\hat{X}$ as in (1), a generator $G$ with parameters $\boldsymbol{\rho}$ can be built following

$$\boldsymbol{\rho}^* = \arg\min_{\boldsymbol{\rho}} \mathbb{E}_{p(X,Y)} \|G_{\boldsymbol{\rho}}(Y) - X\|, \tag{2}$$

where $p(X,Y)$ quantifies the joint distribution of the LR-HR pairs. In testing, given an LR image $Y$, the HR output can then be generated as $\hat{X} = G_{\boldsymbol{\rho}^*}(Y)$.

While the learning-based approaches have become state-of-the-art recently, a bottleneck in successfully applying those methods relates to the fact that the real HR images must be available for training. On the contrary, in the model-based approaches, the real HR images are not always necessary. We consider that, in clinical practice, real HR images are not commonly acquired, and hence it is hard to train such a network. Particularly, in the scenario of reducing MR slice spacing, one may expect to complete the task without referring to the acquisition of HR image set $\{X\}$. Thus, the learning-based methods should be adapted to the lack of real supervision in training, while they are expected to deliver the performance comparable to full supervision.

To this end, we propose the two-stage solution to complete the task of reducing MR image slice spacing, i.e., synthesizing HR images from LR images in Stage I and training $G_{\boldsymbol{\rho}}$ with the supervision of the synthesized HR images in Stage II. Specifically, in Stage I, a VAE network is designed, which synthesizes HR images $\{V(Y)\}$ by interpolating from 2D slices in LR images $\{Y\}$. The set of $\{V(Y)\}$ contains synthesized HR volumes that are realistic in perception, while the anatomical structures can be distorted even in corresponding slices between $Y$ and $V(Y)$. In Stage II, the synthesized HR images $\{V(Y)\}$ are degraded by down-sampling to produce the paired LR images $\{T \cdot V(Y)\}$. We further adopt state-of-the-art deep-learning-based super-resolution method to model the LR-HR mapping upon the synthesized image pairs following

$$\boldsymbol{\rho}^* = \arg\min_{\boldsymbol{\rho}} \mathbb{E}_{p(Y)} \| G_{\boldsymbol{\rho}}(T \cdot V(Y)) - V(Y) \|. \tag{3}$$

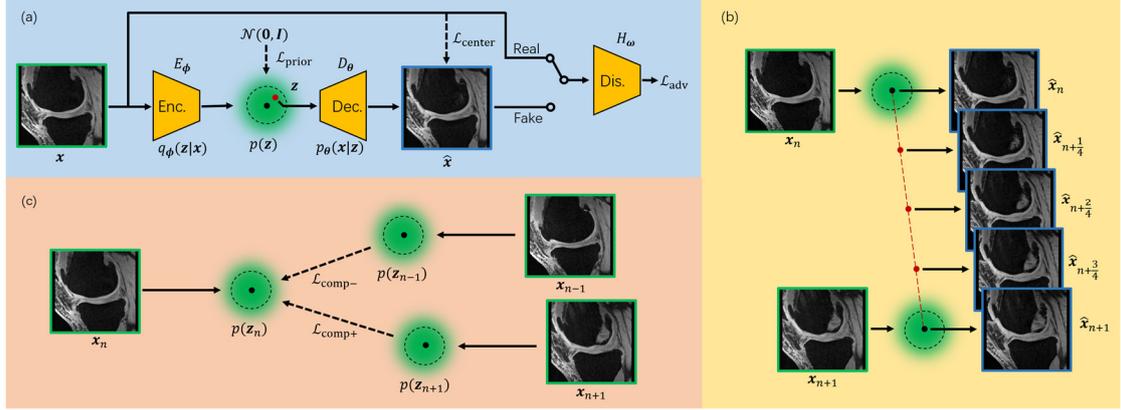

*Fig. 3. Training and testing of VAE-based image up-sampling. (a) In training, a VAE-GAN is optimized over LR image slices in acquisition planes. (b) In testing, neighboring LR slices are interpolated in the latent representation space, and then processed through the decoder to produce many in-between slices. (c) To consider the coherence of anatomical structures in 3D, we propose a novel loss $\mathcal{L}_{comp}$ to enforce slice-wise compactness in the latent space.*

Comparing (3) to (2), our method involves no real HR images $\{X\}$ in training $G_{\boldsymbol{\rho}}$. With only LR images $\{Y\}$ needed, the proposed method belongs to the category of self super-resolution and fits many real medical imaging scenarios flexibly. Meanwhile, our method adopts the full supervision framework of deep-learning-based super-resolution, as the training of the generator is supervised from the synthesized HR images. In this way, we combine both advantages of self super-resolution and supervised super-resolution. We can thus deliver satisfactory super-resolution quality and reduce the performance gap even though no real HR supervision is available to our method.

## 2.2 Synthesis of HR Images

We propose to synthesize HR images from LR images by interpolating in a compactly distributed latent space (cf. Stage I in Fig. 2(b)). The synthesized HR images must appear as realistic as possible, otherwise, the supervision will not deliver satisfactory results in training the super-resolution generator in Stage II. To achieve this, we particularly adopt the encoder-decoder architecture in Fig. 3(a). The VAE encoder ($E_\phi$) maps the 2D slice into the latent space, and the decoder ($D_\theta$) expects to fully reconstruct the slice as the output. With the trained VAE, we can then input two neighboring slices into the encoder and conduct linear interpolation between them in the latent space. The interpolated representations in the latent space are reconstructed by the decoder, which results in missing slices of the original LR image.

### 2.2.1 Variational auto-encoding of 2D slices

Given a 2D slice $x$, we have the variational approximation $q_\phi(z|x)$ with the parameters $\phi$ to encode $x$ as the latent representation $z$. A generative model $p_\theta(x|z)$ with the parameters $\theta$ can decode $z$ and reconstruct the slice $x$. By doing this, the VAE network aims to maximize the evidence lower bound (ELBO) [26] with the following cost function

$$\mathcal{L}_{\text{VAE}} = \underbrace{\underbrace{KL\big(q_\phi(z|x)\|p(z)\big)}_{\mathcal{L}_{\text{prior}}} - \underbrace{\mathbb{E}_{q_\phi(z|x)}\log p_\theta(x|z)}_{\mathcal{L}_{\text{NLL}}}}_{-\text{ELBO}}. \qquad (4)$$

The first term $\mathcal{L}_{\text{prior}}$ in (4) is a regularization term upon the distribution of $z$ in the latent space. The latent representations of individual slices should compactly distribute, in order to facilitate linear interpolation among them. Therefore, we set the prior to be a multi-variate Gaussian, i.e., $p(z) = \mathcal{N}(\mathbf{0}, \mathbf{I})$. The second term $\mathcal{L}_{\text{NLL}}$ penalizes the negative-likelihood (NLL) of $x$ reconstructed with $\phi$ and $\theta$, which encourages the encoder-decoder to transmit comprehensive dataflow for all slices such that $x$ can be fully reconstructed.

The encoder $E_\phi$ and the decoder $D_\theta$ are implemented by deep learning. For probabilistic approximation of the latent representation, VAE usually takes the Gaussian mean-field variational family [27,28], i.e., $q_\phi(z|x) = \mathcal{N}(\mu_x, \sigma_x^2 I)$, while the tuple $(\mu_x, \log \sigma_x^2)$ can be treated to describe the representation of $x$ encoded by $E_\phi$. In this way, $\mathcal{L}_{\text{prior}}$ can be calculated as

$$\mathcal{L}_{\text{prior}} = \text{KL}\big(\mathcal{N}(\boldsymbol{\mu}_x, \boldsymbol{\sigma}_x^2 \boldsymbol{I}) \| \mathcal{N}(\boldsymbol{0}, \boldsymbol{I})\big), \tag{5}$$

where KL(· ∥ ·) computes the Kullback-Leibler divergence.

It is critical for VAE to reconstruct realistic images without degrading their perceptive quality. Kingma *et al.* [26] employed multi-variate Gaussian distribution to model gray-scale images, which may produce blurry outputs though. Larsen *et al.* [29] brought up the idea of generative adversarial network (GAN) [30] and proposed the hybrid VAE-GAN model, which has the pixel-wise loss over discriminator feature maps [31] and the style loss [32]. In this work, we also adopt the idea of VAE-GAN when designing the likelihood term $\mathcal{L}_{\text{NLL}}$. That is, in addition to the fidelity loss $\mathcal{L}_{\text{center}} = \|\boldsymbol{x} - \hat{\boldsymbol{x}}\|$, we train a discriminator $H_{\boldsymbol{\omega}}$ to measure the dissimilarity between the synthesized and real slices [33]. Thus, we have

$$\gamma \mathcal{L}_{\text{NLL}} = \alpha \mathcal{L}_{\text{center}} + \beta \mathcal{L}_{\text{adv}}, \tag{6}$$

$$\text{and} \quad \mathcal{L}_{\text{adv}} = -H_{\boldsymbol{\omega}}(\hat{\boldsymbol{x}}) + H_{\boldsymbol{\omega}}(\boldsymbol{x}), \tag{7}$$

where $\alpha$, $\beta$, and $\gamma$ are non-negative scalars to balance different loss terms.

Notice that in training the VAE network, the latent representation is encoded to quantify $q_{\boldsymbol{\phi}}(\boldsymbol{z}|\boldsymbol{x})$, from which an instance of $\boldsymbol{z}$ will be drawn stochastically. The instance is passed to the decoder, for reconstructing the input slice and evaluating the loss. In testing, however, the instance of $\boldsymbol{z}$ will be directly set as the expectation of $q_{\boldsymbol{\phi}}(\boldsymbol{z}|\boldsymbol{x})$, or $\boldsymbol{z}_x = \boldsymbol{\mu}_x$.

### 2.2.2 Interpolating neighboring slices

It is known that simple arithmetic operation in the image space, e.g., to average two slices even though they are very similar, could degrade the image quality and get blurry output. However, given a valid representation in the latent space learned by VAE, the decoder will be able to reconstruct a high-quality 2D slice. Therefore, we encode two consecutive slices from a 3D volume of large spacing, i.e., $\boldsymbol{x}_n$ and $\boldsymbol{x}_{n+1}$, and perform linear interpolation between their representations in the latent space. After processing these linearly interpolated representations through the decoder, more slices between $\boldsymbol{x}_n$ and $\boldsymbol{x}_{n+1}$ can then be acquired. Specifically, we compute $\boldsymbol{\mu}_n$ and $\boldsymbol{\mu}_{n+1}$ for the slices $\boldsymbol{x}_n$ and $\boldsymbol{x}_{n+1}$. Then we can generate new instances in the representation space as $\hat{\boldsymbol{z}}_{n+\xi} = (1-\xi)\boldsymbol{\mu}_n + \xi\boldsymbol{\mu}_{n+1}$, where $\xi$ is a real scalar between 0 and 1. The decoder thus outputs the slices $\{\hat{\boldsymbol{x}}_{n+\xi}\}$, which is stacked into the 3D volume. An example of the above process is shown in Fig. 3(b), where the spacing between $\boldsymbol{x}_n$ and $\boldsymbol{x}_{n+1}$ is reduced to a quarter after slice interpolation.

Although VAE is capable of synthesizing realistic slices here, the output is not always reasonable after stacking 2D slices back into a 3D volume. The reason is because the encoding-decoding is learned in 2D manner. The network may have ignored intrinsic coherence of anatomical structures in 3D space. To this end, we further propose to enforce slice-wise compactness in learning the latent representation space. That is, regarding the training slice $\boldsymbol{x}_n$, we consider three consecutive slices (i.e., $\boldsymbol{x}_{n-1}$, $\boldsymbol{x}_n$, and $\boldsymbol{x}_{n+1}$) as a whole. The three slices are required to be close with each other in the latent space (cf. Fig. 3(c)), implying that they have similar latent representations:

$$\mathcal{L}_{\text{comp}} = \underbrace{\text{KL}(p(\boldsymbol{z}_n|\boldsymbol{x}_n)\|p(\boldsymbol{z}_{n+1}|\boldsymbol{x}_{n+1}))}_{\mathcal{L}_{\text{comp+}}} + \underbrace{\text{KL}(p(\boldsymbol{z}_n|\boldsymbol{x}_n)\|p(\boldsymbol{z}_{n-1}|\boldsymbol{x}_{n-1}))}_{\mathcal{L}_{\text{comp-}}}. \quad (8)$$

The overall loss in training the VAE network then becomes

$$\mathcal{L}_{\text{VAE}}^{\text{overall}} = \alpha\mathcal{L}_{\text{center}} + \beta\mathcal{L}_{\text{adv}} + \gamma\mathcal{L}_{\text{prior}} + \eta\mathcal{L}_{\text{comp}}, \quad (9)$$

where $\eta$ is another non-negative scalar. Our proposed framework for synthesizing realistic 3D HR images consists of the probabilistic encoder $E_{\boldsymbol{\phi}}$, the decoder $D_{\boldsymbol{\theta}}$ and the discriminator $H_{\boldsymbol{\omega}}$. During training, $E_{\boldsymbol{\phi}}$ is trained with $\beta = 0$ following the practice of Larsen *et al.* [29], while $D_{\boldsymbol{\theta}}$ and $H_{\boldsymbol{\omega}}$ are optimized in an adversarial manner, i.e.,

$$\boldsymbol{\phi}^* = \arg\min_{\boldsymbol{\phi}} \alpha\mathcal{L}_{\text{center}} + \gamma\mathcal{L}_{\text{prior}} + \eta\mathcal{L}_{\text{comp}}, \quad (10)$$

$$\text{and} \quad \boldsymbol{\theta}^*, \boldsymbol{\omega}^* = \arg\min_{\boldsymbol{\theta}}\max_{\boldsymbol{\omega}} \alpha\mathcal{L}_{\text{center}} + \beta\mathcal{L}_{\text{adv}} + \gamma\mathcal{L}_{\text{prior}} + \eta\mathcal{L}_{\text{comp}}. \quad (11)$$

## 2.3 SUPER-RESOLUTION GENERATOR

With the VAE tool $V(\cdot)$, we are capable of synthesizing many HR images from LR inputs only, i.e., $X' = V(Y)$ for each $Y$. Further, in Stage II, we train a super-resolution generator $G_{\boldsymbol{\rho}}$ by using the synthesized HR images for supervision. First, we manually down-sample each synthesized HR image $X'$ to the corresponding LR version $Y'$ using $Y' = T \cdot X'$, with $T$ modeling the degrading processing. In our case, since we only intend to reduce the slice spacing without modifying slicing thickness, $T$ simply takes one slice from $K$ consecutive slices, where $K$ is the down-sampling ratio. Then, we train the generator to estimate HR image from its LR version, over these synthesized LR-HR pairs.

The generator for super-resolution can follow state-of-the-art solutions. In particular, we investigate and compare two generators in this paper, i.e., DCSRN [34] and DeepResolve [17]. DCSRN is primarily designed from DenseNet [19] for the purpose of brain MR image super-resolution; whereas DeepResolve is developed from ResNet [35] and can be considered as the state-of-the-art tool for knee MR image super-resolution.

The generator $G$ is optimized with the reconstruction loss $\mathcal{L}_{\text{rec}}$ over the pairs of the synthesized LR $Y'$ and HR $X'$ images, which is in line with (2):

$$\boldsymbol{\rho}^* = \arg\min_{\boldsymbol{\rho}} \mathbb{E}_{p(X',Y')} \underbrace{\left\| G_{\boldsymbol{\rho}}(Y') - X' \right\|}_{\mathcal{L}_{\text{rec}}}. \tag{12}$$

In our implementation, L1-norm is adopted for the voxel-wise fidelity loss $\|\cdot\|$. Limited by GPU memory, we crop 3D image patches to train the generator. In the testing stage, $G_{\boldsymbol{\rho}}$ can generalize well to real LR image. For each LR image $Y$, its corresponding HR version is directly estimated as $\hat{X} = G_{\boldsymbol{\rho}^*}(Y)$.

## 2.4 SUMMARY

The full pipeline of our proposed unsupervised, two-staged super-resolution framework is summarized in Algorithm 1.

---
**Algorithm 1. Reducing MRI Slice Spacing without Ground-Truth**

---
**Input:** LR volumes $\{Y\}$ of large slice spacing, where $Y = \{x_1, x_2, ..., x_N\}$.
**Output:** Estimated HR volumes $\{\hat{X}\}$ of slice spacing reduced to $\frac{1}{K}$, where $\hat{X} = \left\{\hat{x}_1, \hat{x}_{1+\frac{1}{K}}, ..., \hat{x}_N\right\}$.

**Stage I:** Produce synthesized HR volumes $\{X'\}$ from LR volumes $\{Y\}$ only.
    **Stage I-1:** $\boldsymbol{\phi}^* = \arg\min_{\boldsymbol{\phi}} \mathbb{E}_{p(x_n)} \alpha \mathcal{L}_{\text{center}} + \gamma \mathcal{L}_{\text{prior}} + \eta \mathcal{L}_{\text{comp}}$,
    and $\boldsymbol{\theta}^*, \boldsymbol{\omega}^* = \arg\min_{\boldsymbol{\theta}} \max_{\boldsymbol{\omega}} \mathbb{E}_{p(x_n)} \alpha \mathcal{L}_{\text{center}} + \beta \mathcal{L}_{\text{adv}} + \gamma \mathcal{L}_{\text{prior}} + \eta \mathcal{L}_{\text{comp}}$.
    // Optimize 2D VAE-GAN ($E_{\boldsymbol{\phi}}$, $D_{\boldsymbol{\theta}}$, and $H_{\boldsymbol{\omega}}$) on 2D slices $\{x_n\}$ from LR volumes $\{Y\}$.
    **Stage I-2:** Synthesize HR volumes $\{X' \mid X' = V(Y)\}$ from LR volumes $\{Y\}$ with VAE-based up-sampling tool $V(\cdot)$.
        (a) $z_n = \boldsymbol{\mu}_n$ where $(\boldsymbol{\mu}_n, \boldsymbol{\sigma}_n) = E_{\boldsymbol{\phi}^*}(x_n)$.
        // Map 2D slices $\{x_n\}$ extracted from LR volumes $\{Y\}$ to latent representations
        (b) $\hat{z}_{n+\xi} = (1-\xi) z_n + \xi z_{n+1}$ where $\xi = \left\{\frac{0}{K}, \frac{1}{K}, ..., \frac{K}{K}\right\}$.
        // Linearly interpolate representations of neighboring slices in LR volumes.
        (c) $x'_{n+\xi} = D_{\boldsymbol{\theta}^*}(\hat{z}_{n+\xi})$ and $X' = \left\{x'_1, x'_{1+\frac{1}{K}}, ..., x'_N\right\}$.
        // Map interpolated representations $\{\hat{z}_{n+\xi}\}$ back to 2D slices $\{x'_{n+\xi}\}$, and compose them to synthesized HR volumes $\{X'\}$.
**Stage II:** Optimize 3D super-resolution network $G_{\boldsymbol{\rho}}$ with synthesized HR volumes $\{X'\}$.
    **Stage II-1:** $Y' = \{x'_1, x'_2, ..., x'_N\}$.
    // Degrade synthesized HR volumes $\{X'\}$ to synthesized LR volumes $\{Y'\}$.
    **Stage II-2:** $\boldsymbol{\rho}^* = \arg\min_{\boldsymbol{\rho}} \mathbb{E}_{p(X',Y')} \|G_{\boldsymbol{\rho}}(Y') - X'\|$.
    // Optimize a 3D super-resolution generator $G_{\boldsymbol{\rho}}$ on synthesized LR-HR pairs $\{Y', X'\}$.

**Testing:**
    $\hat{X} = G_{\boldsymbol{\rho}^*}(Y)$
    // Estimate HR volumes $\{X\}$ with super-resolution generator and input LR volumes $\{Y\}$.

---

# 3 EXPERIMENTS

We have applied our proposed framework to knee MR images to verify its performance in reducing slice spacing. There are two datasets used in this work – a publicly accessible dataset of *Segmentation of Knee Images 2010* (SKI10) [36] and a private dataset of clinical knee MR acquisitions from our institution. We first introduce experimental settings in Section 3.1. Then, detailed performance comparisons are provided in Section 3.2 for SKI10 and Section 3.3 for the private dataset. Finally, we have conducted ablation studies for the critical VAE-based HR image synthesis in Section 3.4.

## 3.1 EXPERIMENTAL SETTINGS

Our framework is implemented with PyTorch [37] and optimized with graphics processing unit (GPU) cards. The 2D slices $x$ are with size 256×256 in training, and the size of latent representation $z$ is 512. The encoder $E_\phi$ and decoder $D_\theta$ use activation-normalization-convolution blocks and fully connected layers for feature extraction, while average pooling layers and nearest-neighbor interpolation are adopted to change feature map size. The discriminator $H_\omega$ adopts the idea from PatchGAN [38] and does not use any fully connected layer. Other than VAE-GAN, two 3D super-resolution generators, DCSRN [34] and DeepResolve [17], are considered to integrate with our framework. To train the encoder, decoder, and discriminator, Adam optimizer [39] is applied with default hyper-parameters (0.9,0.999) and a learning rate of 2e-4. The batch size is limited by GPU memory and is set to 16. The networks are trained for a total of 2e6 iterations, including a warm-up for 1e5 iterations with the hyper-parameters $(\alpha, \beta, \gamma, \eta)$=(10000, 0, 1, 0), then 8e5 iterations with $(\alpha, \beta, \gamma, \eta)$=(100, 1, 0.05, 0), and finally 1.2e6 iterations with all losses enabled, i.e., $(\alpha, \beta, \gamma, \eta)$=(100, 1, 0.05, 0.02). The generator $G_\rho$ is trained following the configurations in [34] and [17], respectively.

## 3.2 COMPARISONS TO EXISTING METHODS ON SKI10

We have collected a total 150 knee joint MR images from the SKI10 dataset. Among them, 100 images are used for training all networks, and 50 images are left-out for test. All images are in the sagittal plane with the voxel spacing of 0.4mm×0.4mm×1mm. The acquisitions in SKI10 were scanned using machines from major vendors at over 80 different centers, with more imaging details in [36]. In our experiment, we verify the performance to reduce slice spacing by four folds, or $K$=4. Quantitative comparisons are listed in Table 1. The mean and standard deviation (SD) of two widely used metrics, i.e., peak signal-to-noise ratio (PSNR) and structural similarity index (SSIM) [40], are used to measure the performance of different methods. Also, we have provided visual comparisons for both in-plane (Fig. 4) and through-plane views (Fig. 5).

**Table 1**
Quantitative comparison of different super-resolution methods.

| Method | PSNR | | SSIM | |
|---|---|---|---|---|
| | Mean | SD | Mean | SD |
| Trilinear Up-Sample | 27.152 | 2.153 | 0.797 | 0.042 |
| EDSSR [18] | 28.070 | 2.281 | 0.823 | 0.034 |
| VAE-Based Up-Sample | 18.995 | 2.003 | 0.358 | 0.074 |
| DCSRN Only [29] | 28.910 | 2.312 | 0.850 | 0.036 |
| Proposed with DCSRN | 28.470 | 2.319 | 0.835 | 0.040 |
| DeepResolve Only [12] | 31.005 | 2.435 | 0.891 | 0.033 |
| Proposed with DeepResolve | **30.058** | **2.377** | **0.870** | **0.035** |

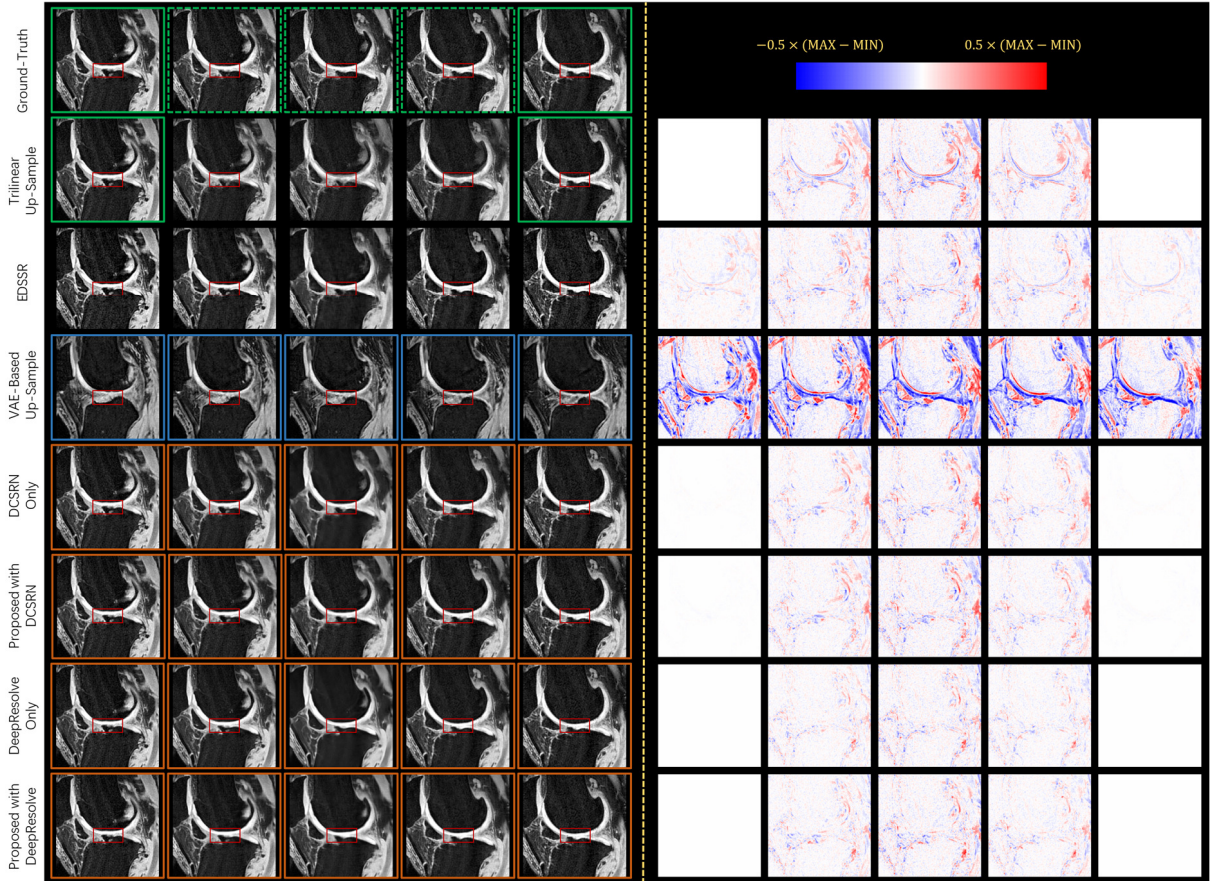

*Fig. 4. Visual comparison between different super-resolution algorithms from sagittal (in-plane) view. From left to right there are five consecutive slices in each row. The first and last slices of the ground-truth (in solid green boxes) are observed in LR image and input for super-resolution, while the outputs can be compared with the middle three ground-truth slices (in dashed green boxes). The error maps of the super-resolution results with respect to the ground-truth are shown in the right of the figure. Rows from top to bottom correspond to the ground-truth, trilinear up-sampling, EDSSR [18], VAE-based up-sampling (Stage I in the proposed method), DCSRN [29] trained with real HR supervision (DCSRN only), our proposed framework with DCSRN in Stage II, DeepResolve [12] trained with real HR supervision (DeepResolve Only), and our proposed framework with DeepResolve in Stage II.*

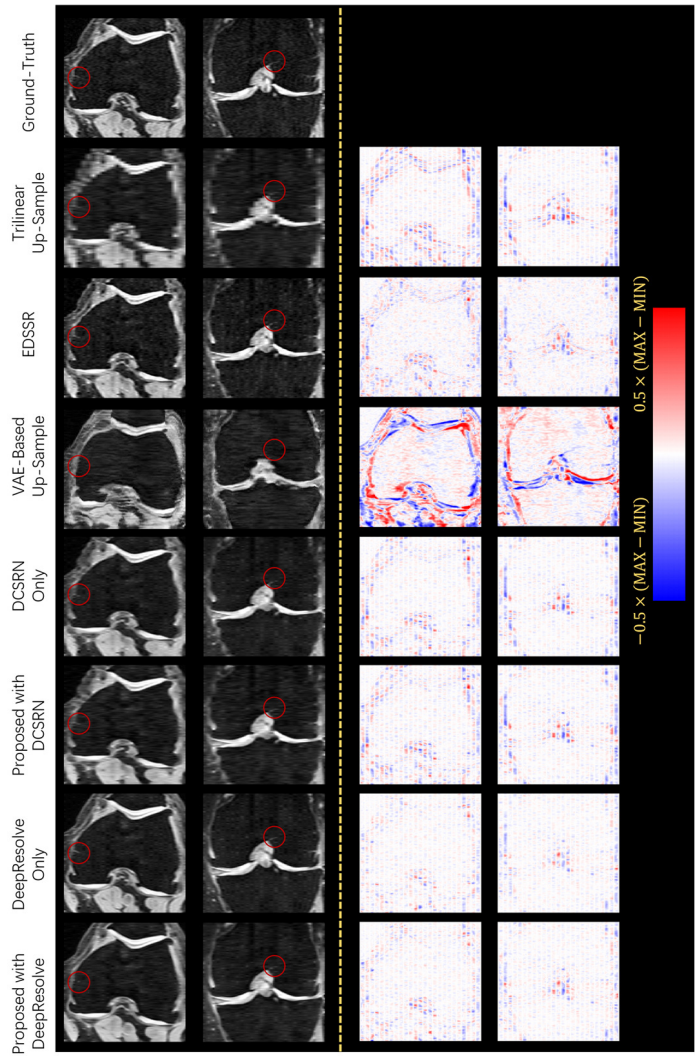

*Fig. 5. Visual comparison between different super-resolution algorithms from through-plane views. The first/second column accounts for the axial/coronal slices, respectively, with the corresponding error maps to the ground-truth on the right. The subject and the methods compared here are the same with those used in Fig. 4.*

In Fig. 4, we select a test image to reduce its slice spacing. For every five consecutive slices in the HR image, we only input the first and the last slices (outlined by solid green boxes in the figure) to the super-resolution generator. The three slices in the middle can then be treated as the ground-truth (in dashed green boxes) for quantitative evaluation. One may use traditional trilinear up-sampling to interpolate the three middle slices from the first/last inputs. However, the interpolation inevitably deteriorates image quality, resulting in blurred cartilages such as those highlighted by red rectangles in Fig. 4.

Alternatively, EDSSR [23] provides a learning-based solution to reduce slice spacing. This method trains a 2D neural network with sagittal slices and then applies it to reconstruct HR coronal and axial views. In this way, EDSSR has produced plausible HR images compared to simple interpolation in the original image space. The measurements in PSNR and SSIM are also improved accordingly as in Table 1.

The proposed VAE network in Stage I of our framework is capable of up-sampling images. In particular, the input slices are encoded first, interpolated in the latent space, and then decoded to produce the in-between slices. Thanks to the encoding-decoding capability and the powerful discriminator in our network, we can interpolate perceptively realistic slices as in Fig. 4 (highlighted by blue boxes). For example, in the red rectangles, the boundaries of the cartilages are sharp and smooth. However, we observe extremely high voxel-to-voxel errors (also corresponding to low PSNR/SSIM in Table 1) when comparing the VAE up-sampled images with the ground-truth. The reason is that the VAE-based up-sampling has altered and deformed anatomical structures; thus, the metrics of PSNR/SSIM cannot function well due to the lack of voxel-to-voxel correspondences between the up-sampled images and the ground-truth. In general, we conclude that it is not suitable for the VAE network in Stage I to complete the task of reducing slice spacing alone.

To complete Stage II of our proposed framework, we adopt existing fully supervised methods, i.e., DCSRN and DeepResolve, to train the super-resolution generators. When integrated with DCSRN, our method achieves PSNR=28.470 and SSIM=0.835; when integrated with DeepResolve, our method can achieve PSNR=30.058 and SSIM=0.870. The above measures are all higher than the scores achieved by EDSSR (PSNR=28.070, SSIM=0.823). The quantitative measures are also in line with the visual comparisons in Fig. 4 and Fig. 5, where the results produced by the proposed framework are always closer to the ground-truth regardless of the adopted super-resolution generator. The error maps show that, especially at the places of the (first/last) input slices, our method performs much better than EDSSR. The reason is that our method can fully preserve the input sagittal slices and estimate the in-between slices; in EDSSR, however, HR coronal/axial views are generated for fusion of the 3D volume, implying inevitable loss to the in-plane quality in sagittal views. As a summary, the above results demonstrate superior performance of our framework compared to EDSSR, which is the state-of-the-art self super-resolution method.

The performance of our framework is still bounded by the fully supervised generator adopted in Stage II. However, our framework relying on LR images only can deliver high super-resolution capability that is near the case when LR-HR supervision is available. Assuming that the synthesized HR images are indistinguishable to real HR images, theoretically our method can be very close to DCSRN/DeepResolve that is trained on real LR-HR pairs. Although there are still performance gaps in Table 1 when comparing "Proposed with DCSRN" to "DCSRN Only" and "Proposed with DeepResolve" to "DeepResolve Only", the visual differences are relatively small referring to the error maps in Fig. 4 and Fig. 5. Meanwhile, we observe that DeepResolve tends to be a better generator to connect with our framework, which will be adopted in subsequent experiments.

It is worth noting that, for all training involved in Table 1, we only use 100 LR-HR pairs from the SKI10 dataset. The rest 50 pairs are used as the ground-truth for evaluation and isolated from training. In fact, our proposed framework requires no HR supervision, implying that we can count in the left-out ground-truth data to further enlarge the training dataset. In this experiment, however, we stick to the same training-testing data split for fair comparisons of all methods, even though our framework could utilize more training data for potentially better performance.

## 3.3 COMPARISONS TO EXISTING METHODS ON PRIVATE CLINICAL DATASET

We have further applied our proposed super-resolution framework to real knee MR images that are acquired for clinical routines. We collect a total of 2627 subjects and center-crop all sagittal slices to the size 256×256 with the resolution 0.4mm×0.4mm. The slice spacing is 3.3mm or 4.5mm. Our goal is still to reduce the slice spacing by four folds ($K$=4). Different from SKI10, all images can be used to train our framework while we have no ground-truth for this dataset.

Visual comparisons of different super-resolution methods are provided in Fig. 6. In the first row of the figure, we have the first and last slices as the inputs, while the three slices in the middle can be acquired through trilinear interpolation. However, the interpolation-based up-sampling brings in many artifacts, such as the incorrectly bifurcating cartilage structures in red rectangles across slices. In the second row, the VAE-based synthesis can produce much more realistic images. However, the anatomical structured are again altered even in the places where the input slices are available (i.e., the first/last column in the figure, with an example highlighted in red circles). That is, the VAE-based synthesis here can hardly attain the goal of super-resolution. On the contrary, by adopting the generator of DeepResolve, our method can restore the unobserved slices with plausible image quality, while the observed slices in LR inputs can still be well preserved.

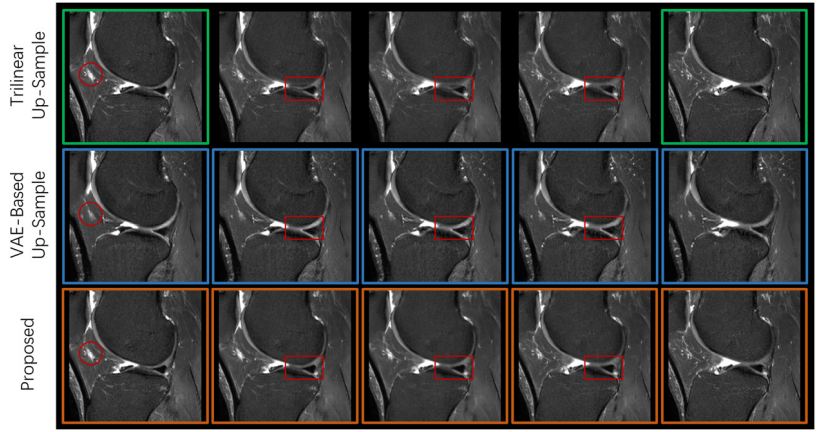

*Fig. 6. Visual comparison of different methods on the private knee MR image dataset. With two slices in LR acquisition as the inputs (in green boxes), we acquire HR slices (K=4) by trilinear up-sampling, the VAE-base up-sampling, and the proposed framework.*

## 3.4 ABLATION STUDIES OF THE VAE-BASED SYNTHESIS

It is critical that we can synthesize realistic HR images in the proposed framework. To this end, we have designed the VAE tool in Stage I, which is powered by the novel loss in (9). Only if the individual 2D slices are compactly encoded in the latent space, the interpolation and subsequent decoding can yield plausible 3D HR volumes.

### 3.4.1 The Role of $\mathcal{L}_{adv}$

We first analyze the effect of adversarial loss $\mathcal{L}_{adv}$. With the same settings in Section 3.1, two VAEs are trained over the SKI10 dataset for 4e5 iterations after warm-up. One of them is trained without $\mathcal{L}_{adv}$ by setting $\beta$=0. The two VAEs can be used to reconstruct slices from the SKI10 testing dataset. Fig. 7 provides the examples of the slices produced through different VAEs. It is clear that, without $\mathcal{L}_{adv}$, VAE can hardly produce high-quality slices that are able to supervise the training of the super-resolution generator in Stage II.

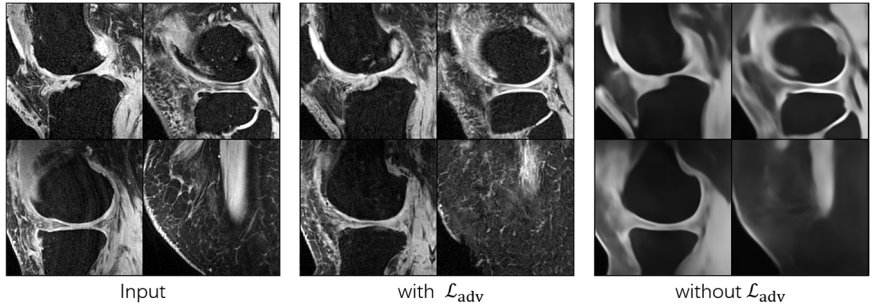

*Fig. 7. Visual comparison of the slices reconstructed by VAEs, which are trained with and without $\mathcal{L}_{adv}$. The four exemplar slices are randomly selected from the SKI10 testing set.*

### 3.4.2 The Role of $\mathcal{L}_{comp}$

The term $\mathcal{L}_{comp}$ encourages the local compactness of the latent representation space for individual slices. Particularly, we train two VAEs with the same settings described in Section 3.1, except that $\eta$ is enabled/disabled. In Fig. 8, an example from the SKI10 testing dataset is used, and the VAE up-sampled slices are extracted and rendered from the coronal view. Without $\mathcal{L}_{comp}$, the synthesized HR image may introduce unrealistic anatomical structures, e.g., the fluctuating femoral/tibial cartilage interfaces in zoom-in views. On the contrary, with $\mathcal{L}_{\text{comp}}$ enabled, smooth cartilage interfaces that are physiologically reasonable can be acquired. Note that the outputs of VAEs can deviate significantly from the ground-truth in terms of voxel-to-voxel correspondences. However, thanks to the combination of the VAE network in Stage I and the DeepResolve generator in Stage II, our framework can still recover the missing sagittal slices successfully.

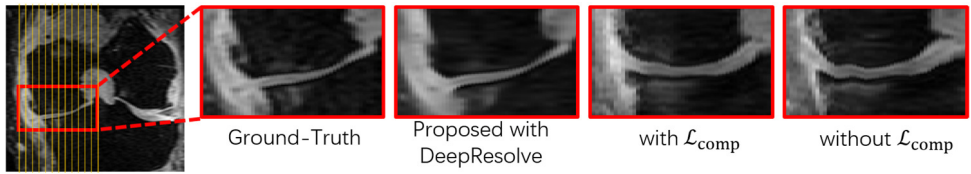

*Fig. 8. Visual comparison in coronal views for the volumes up-sampled by VAEs, which are trained with and without $\mathcal{L}_{comp}$. The yellow grid indicates the positions of the input LR (sagittal) slices. Zoom-in views show that, with $\mathcal{L}_{comp}$, the structures of bone/cartilage interfaces become physiologically reasonable. Note that the outputs of VAEs can deviate from the ground-truth significantly. However, our framework can still recover the missing sagittal slices thanks to the combination of the VAE network in Stage I and the DeepResolve generator in Stage II.*

We further examine the local compactness of the latent space by referring to mutual distances of the encoded slices in the SKI10 testing set. Given the representations $\{\boldsymbol{\mu}_n, \boldsymbol{\mu}_{n+1}\}$ of arbitrary neighboring slices from an LR image, the corresponding HR representations after linear interpolation can be obtained as $\{\boldsymbol{\mu}_n, \cdots, \hat{\boldsymbol{z}}_{n+\xi}, \cdots, \boldsymbol{\mu}_{n+1}\}$, which are represented by the dashed line and the corresponding red dots in Fig. 9. As we have the ground-truth for the missing slices, we can encode the ground-truth slices directly and acquire $\boldsymbol{z}_{n+\xi}$ explicitly (designated by black dots in the figure). If the encoded space is highly linear, which is necessary to our interpolation, the mismatch between $\boldsymbol{z}_{n+\xi}$ and $\hat{\boldsymbol{z}}_{n+\xi}$ should be small. To this end, we compute the distances between the encoded slices in the latent space. Particularly, $\xi$ can be 1/4, 2/4, or 3/4. We thus define three types of distances: $d_1$ for the distance between $\boldsymbol{z}_{n+\xi}$ and $\hat{\boldsymbol{z}}_{n+\xi}$ when $\xi$ is 1/4 or 3/4, $d_2$ for $\xi=2/4$, and $d_3$ for the distance between $\boldsymbol{z}_{n+\xi}$'s of two neighboring slices. As in the figure, with $\mathcal{L}_{\text{comp}}$ enabled for training the VAE network, $d_1$ and $d_2$ both decrease significantly ($p<0.01$, paired $t$-tests), implying that the interpolated representations can better approximate the ground-truth slices in the latent space. Meanwhile, $d_3$ also decreases significantly ($p<0.01$, paired *t*-tests), implying that all slices from the same subject are more locally distributed. The compactness of the latent space thus benefits HR image synthesis and subsequent supervised super-resolution.

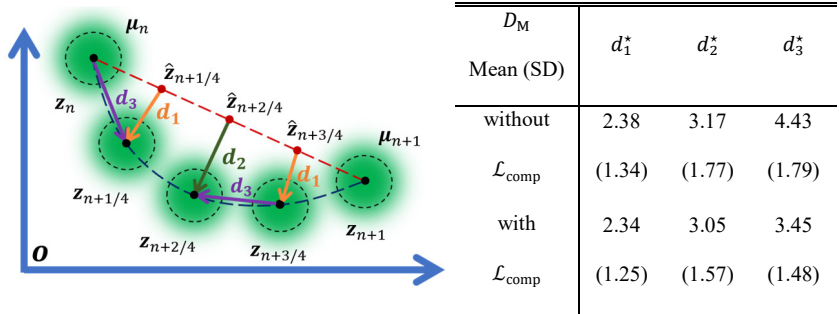

★ Significant difference (with paired $t$-test and $p<0.01$).

Fig. 9. Quantitative comparison between VAEs trained with and without $\mathcal{L}_{comp}$. With $\mathcal{L}_{comp}$, distances between linearly interpolated codes (red dots) and growth-truth (black dots) are significantly smaller ($d_1$ and $d_2$) in the latent space. Also, codes for neighboring ground-truth slices (black dots) are significantly closer to each other ($d_3$) with $\mathcal{L}_{comp}$, indicating they are more locally compact in the latent space. All distances are measured in the SKI10 testing set.

## 4 Discussion and Conclusion

In this work, we have proposed a novel self super-resolution framework to reduce the slice spacing of MR images and this framework consists of HR image synthesis and supervised super-resolution. Supervised super-resolution based on neural-networks delivers leading performance nowadays, however it requires HR images for training. While no HR images are available, which is the common case in various clinical scenarios, we can synthesize HR images from LR inputs effectively. The synthesized HR images can then act as the supervision for a well-designed super-resolution generator. The perceptive quality of the synthesized images is critical to the optimization of the following super-resolution generator, and a novel VAE network synthesizing HR image from LR ones was carefully designed and optimized with strong regularization, such that 2D slices can be encoded compactly. In this way, the linear interpolation of the 2D slices in the latent space can guarantee high-quality 3D volumes after decoding, which significantly outperforms the fuzzy volumes interpolated in the original image space. However, the VAE network cannot guarantee voxel-to-voxel correspondences after encoding-decoding; thus we conclude that the VAE alone may be not suitable to complete task to reducing the slice spacing of medical images. The limitation of VAE motivates us to train the following super-resolution generator, which produces satisfactory results.

The success of our proposed two-stage method generally benefits from the combination of HR image synthesis and supervised super-resolution. Firstly, our proposed framework does not require any real HR images for training, as HR images are not available in many real clinical scenarios. Secondly, without real HR images, our proposed framework performs better than existing self-super-resolution methods for MR volumes. It also achieves visually comparable results with respect to the super-resolution methods that are fully supervised.

There are admittedly several limitations, which will be the future research directions for our work. First, although we have achieved visually comparable results with fully supervised methods, there is still a performance gap between them. We will investigate the ways to further narrow the gap, such as unifying the two stages into a single network for joint learning. Second, the slice spacing reduction ratio K is fixed in the super-resolution generators in Stage II. That is, if we want to apply our framework to a new scenario with a different ratio K, we have to retrain the super-resolution generator used in Stage II. In the future work, we will address this issue to deliver a more flexible super-resolution generator. Finally, our work is very distinct to MR image reconstruction. Although we aim to reduce the slice spacing, our framework cannot add data that are not scanned in MR acquisition. We emphasize that, with our two-staged framework, we can effectively mitigate the large slice spacing and deliver better rendering in 3D. We also note that our framework has a large potential to be applied to may scenarios, which are not restricted to knee or MR. We will investigate its applications in relevant studies in the future.

## ACKNOWLEDGEMENTS

This work was partially supported by the National Key Research and Development Program of China (2018YFC0116400), STCSM (19QC1400600) and the China Scholarship Council.

## REFERENCES

[1]  Y. Jia, A. Gholipour, Z. He, S.K. Warfield, A New Sparse Representation Framework for Reconstruction of an Isotropic High Spatial Resolution MR Volume From Orthogonal

Anisotropic Resolution Scans, IEEE Transactions on Medical Imaging. 36 (2017) 1182–1193. https://doi.org/10.1109/TMI.2017.2656907.

[2]     B. Chen, K. Xiang, Z. Gong, J. Wang, S. Tan, Statistical Iterative CBCT Reconstruction Based on Neural Network, IEEE Transactions on Medical Imaging. 37 (2018) 1511–1521. https://doi.org/10.1109/TMI.2018.2829896.

[3]     F. Shi, J. Cheng, L. Wang, P. Yap, D. Shen, LRTV: MR Image Super-Resolution With Low-Rank and Total Variation Regularizations, IEEE Transactions on Medical Imaging. 34 (2015) 2459–2466. https://doi.org/10.1109/TMI.2015.2437894.

[4]     S. Tourbier, X. Bresson, P. Hagmann, J.-P. Thiran, R. Meuli, M.B. Cuadra, An efficient total variation algorithm for super-resolution in fetal brain MRI with adaptive regularization, NeuroImage. 118 (2015) 584–597. https://doi.org/10.1016/j.neuroimage.2015.06.018.

[5]     J. Hatvani, A. Basarab, J. Tourneret, M. Gyöngy, D. Kouamé, A Tensor Factorization Method for 3-D Super Resolution With Application to Dental CT, IEEE Transactions on Medical Imaging. 38 (2019) 1524–1531. https://doi.org/10.1109/TMI.2018.2883517.

[6]     A.V. Dalca, K.L. Bouman, W.T. Freeman, N.S. Rost, M.R. Sabuncu, P. Golland, Medical Image Imputation From Image Collections, IEEE Transactions on Medical Imaging. 38 (2019) 504–514. https://doi.org/10.1109/TMI.2018.2866692.

[7]     J. Yang, J. Wright, T.S. Huang, Y. Ma, Image Super-Resolution Via Sparse Representation, IEEE Transactions on Image Processing. 19 (2010) 2861–2873. https://doi.org/10.1109/TIP.2010.2050625.

[8]     L. Breiman, Random Forests, Machine Learning. 45 (2001) 5–32. https://doi.org/10.1023/A:1010933404324.

[9]     J. Zhang, L. Zhang, L. Xiang, Y. Shao, G. Wu, X. Zhou, D. Shen, Q. Wang, Brain atlas fusion from high-thickness diagnostic magnetic resonance images by learning-based

super-resolution, Pattern Recognition. 63 (2017) 531–541.

https://doi.org/10.1016/j.patcog.2016.09.019.

[10]     C. Dong, C.C. Loy, K. He, X. Tang, Image Super-Resolution Using Deep Convolutional Networks, IEEE Transactions on Pattern Analysis and Machine Intelligence. 38 (2016) 295–307.

[11]     J. Long, E. Shelhamer, T. Darrell, Fully Convolutional Networks for Semantic Segmentation, in: Proceedings of the IEEE Conference on Computer Vision and Pattern Recognition, 2015: pp. 3431–3440.

[12]     K. Jiang, Z. Wang, P. Yi, J. Jiang, Hierarchical dense recursive network for image super-resolution, Pattern Recognition. 107 (2020) 107475. https://doi.org/10.1016/j.patcog.2020.107475.

[13]     Y. Yang, Y. Qi, Image super-resolution via channel attention and spatial graph convolutional network, Pattern Recognition. 112 (2021) 107798. https://doi.org/10.1016/j.patcog.2020.107798.

[14]     K. Han, Y. Huang, C. Song, L. Wang, T. Tan, Adaptive super-resolution for person re-identification with low-resolution images, Pattern Recognition. (2020) 107682. https://doi.org/10.1016/j.patcog.2020.107682.

[15]     R. Abiantun, F. Juefei-Xu, U. Prabhu, M. Savvides, SSR2: Sparse signal recovery for single-image super-resolution on faces with extreme low resolutions, Pattern Recognition. 90 (2019) 308–324. https://doi.org/10.1016/j.patcog.2019.01.032.

[16]     L. Zhao, H. Bai, J. Liang, B. Zeng, A. Wang, Y. Zhao, Simultaneous color-depth super-resolution with conditional generative adversarial networks, Pattern Recognition. 88 (2019) 356–369. https://doi.org/10.1016/j.patcog.2018.11.028.

[17] A.S. Chaudhari, Z. Fang, F. Kogan, J. Wood, K.J. Stevens, E.K. Gibbons, J.H. Lee, G.E. Gold, B.A. Hargreaves, Super-resolution musculoskeletal MRI using deep learning, Magnetic Resonance in Medicine. 80 (2018) 2139–2154. https://doi.org/10.1002/mrm.27178.

[18] Y. Chen, F. Shi, A.G. Christodoulou, Z. Zhou, Y. Xie, D. Li, Efficient and Accurate MRI Super-Resolution using a Generative Adversarial Network and 3D Multi-Level Densely Connected Network, in: International Conference on Medical Image Computing and Computer-Assisted Intervention, Springer, 2018: pp. 91–99.

[19] G. Huang, Z. Liu, L. Van Der Maaten, K.Q. Weinberger, Densely Connected Convolutional Networks, in: Proceedings of the IEEE Conference on Computer Vision and Pattern Recognition, 2017: pp. 470--4708.

[20] D. Shen, G. Wu, H.-I. Suk, Deep Learning in Medical Image Analysis, Annual Review of Biomedical Engineering. 19 (2017) 221–248. https://doi.org/10.1146/annurev-bioeng-071516-044442.

[21] D. Glasner, S. Bagon, M. Irani, Super-resolution from a single image, in: 2009 IEEE 12th International Conference on Computer Vision, 2009: pp. 349–356. https://doi.org/10.1109/ICCV.2009.5459271.

[22] J.-B. Huang, A. Singh, N. Ahuja, Single Image Super-Resolution From Transformed Self-Exemplars, in: 2015: pp. 5197–5206. https://www.cv-foundation.org/openaccess/content_cvpr_2015/html/Huang_Single_Image_Super-Resolution_2015_CVPR_paper.html (accessed June 9, 2019).

[23] C. Zhao, A. Carass, B.E. Dewey, J.L. Prince, Self super-resolution for magnetic resonance images using deep networks, in: 2018 IEEE 15th International Symposium on Biomedical Imaging (ISBI 2018), 2018: pp. 365–368. https://doi.org/10.1109/ISBI.2018.8363594.

[24] A. Jog, A. Carass, J.L. Prince, Self Super-Resolution for Magnetic Resonance Images, in: S. Ourselin, L. Joskowicz, M.R. Sabuncu, G. Unal, W. Wells (Eds.), Medical Image Computing and Computer-Assisted Intervention -- MICCAI 2016, Springer International Publishing, 2016: pp. 553–560.

[25] B. Lim, S. Son, H. Kim, S. Nah, K.M. Lee, Enhanced Deep Residual Networks for Single Image Super-Resolution, in: Proceedings of the IEEE Conference on Computer Vision and Pattern Recognition Workshops, 2017: pp. 136–144.

[26] D.P. Kingma, M. Welling, Auto-Encoding Variational Bayes, ArXiv:1312.6114 [Cs, Stat]. (2013).

[27] D.M. Blei, A. Kucukelbir, J.D. McAuliffe, Variational Inference: A Review for Statisticians, Journal of the American Statistical Association. 112 (2017) 859–877. https://doi.org/10.1080/01621459.2017.1285773.

[28] M.I. Jordan, Z. Ghahramani, T.S. Jaakkola, L.K. Saul, An Introduction to Variational Methods for Graphical Models, Machine Learning. 37 (1999) 183–233. https://doi.org/10.1023/A:1007665907178.

[29] A.B.L. Larsen, S.K. Sønderby, H. Larochelle, O. Winther, Autoencoding beyond pixels using a learned similarity metric, in: International Conference on Machine Learning, 2016: pp. 1558–1566.

[30] I.J. Goodfellow, J. Pouget-Abadie, M. Mirza, B. Xu, D. Warde-Farley, S. Ozair, A. Courville, Y. Bengio, Generative Adversarial Networks, in: Advances in Neural Information Processing Systems, 2014: pp. 2672–2680.

[31] J. Johnson, A. Alahi, L. Fei-Fei, Perceptual Losses for Real-Time Style Transfer and Super-Resolution, in: European Conference on Computer Vision, Springer, 2016: pp. 694–711.

[32] C. Ledig, L. Theis, F. Huszar, J. Caballero, A. Cunningham, A. Acosta, A. Aitken, A. Tejani, J. Totz, Z. Wang, W. Shi, Photo-Realistic Single Image Super-Resolution Using a Generative Adversarial Network, in: Proceedings of the IEEE Conference on Computer Vision and Pattern Recognition, 2017: pp. 4681–4690.

[33] M. Arjovsky, S. Chintala, L. Bottou, Wasserstein Generative Adversarial Networks, in: International Conference on Machine Learning, 2017: pp. 214–223.

[34] Y. Chen, Y. Xie, Z. Zhou, F. Shi, A.G. Christodoulou, D. Li, Brain MRI super resolution using 3D deep densely connected neural networks, in: 2018 IEEE 15th International Symposium on Biomedical Imaging (ISBI 2018), 2018: pp. 739–742. https://doi.org/10.1109/ISBI.2018.8363679.

[35] K. He, X. Zhang, S. Ren, J. Sun, Deep Residual Learning for Image Recognition, in: Proceedings of the IEEE Conference on Computer Vision and Pattern Recognition, 2016: pp. 770–778.

[36] T. Heimann, B.J. Morrison, M.A. Styner, M. Niethammer, S. Warfield, Segmentation of knee images: a grand challenge, in: Proc. MICCAI Workshop on Medical Image Analysis for the Clinic, 2010: pp. 207–214.

[37] A. Paszke, S. Gross, S. Chintala, G. Chanan, E. Yang, Z. DeVito, Z. Lin, A. Desmaison, L. Antiga, A. Lerer, Automatic differentiation in PyTorch, (2017). https://openreview.net/forum?id=BJJsrmfCZ.

[38] P. Isola, J.-Y. Zhu, T. Zhou, A.A. Efros, Image-to-Image Translation with Conditional Adversarial Networks, in: Proceedings of the IEEE Conference on Computer Vision and Pattern Recognition, 2017: pp. 1125–1134.

[39] D.P. Kingma, J. Ba, Adam: A Method for Stochastic Optimization, ArXiv:1412.6980 [Cs]. (2014).
30

[40]  Z. Wang, E.P. Simoncelli, A.C. Bovik, Multiscale structural similarity for image quality assessment, in: The Thrity-Seventh Asilomar Conference on Signals, Systems Computers, 2003, 2003: pp. 1398-1402 Vol.2. https://doi.org/10.1109/ACSSC.2003.1292216.